\documentstyle[prb,aps,preprint]{revtex}
\begin{document}
\draft
\title{\large{\bf Hidden dimers and the matrix maps: Fibonacci 
chains re-visited}} 
\vskip .2in
\author{Samar Chattopadhyay$^1$ and Arunava Chakrabarti$^2$}
\vskip .2in
\address{
$^1$Department of Physics, Hooghly Mohsin College, Chinsurah, 
West Bengal 712 101, India.\\
$^2$Department of Physics, 
University of Kalyani,
Kalyani, West Bengal 741 235, India.}
\maketitle
\vskip .3in
\begin{abstract}
The existence of cycles of the matrix maps in Fibonacci class of
lattices is well established. We show that such cycles are intimately
connected with the presence of interesting positional correlations among the
constituent `atoms' in a one dimensional quasiperiodic lattice. We
particularly address the 
transfer model of the classic golden mean 
Fibonacci chain where a six cycle of the
full matrix map exists at the centre of the spectrum [Kohmoto et al, 
Phys. Rev. {\bf B 35}, 1020 (1987)], and for which no
simple physical picture has so far been provided, to the best of our knowledge.
In addition, we show that our prescription leads to a determination of
other energy values 
for a mixed model of the Fibonacci chain, 
for which the full matrix map may have similar cyclic
behaviour. Apart from the standard transfer-model of a golden mean
Fibonacci chain, 
we address a variant of it and the silver mean lattice, where the
existence of four cycles of the matrix map is already known to exist. 
The underlying positional
correlations for all such cases are discussed in details.
\end{abstract}
\nopagebreak
\newpage
\noindent
One dimensional quasiperiodic lattices have been the subject of 
extensive research in the past several years, 
and we have already gathered a wealth of knowledge about these systems
[1-11,13-16, and references therein]. 
Theoretical studies of the spectral properties of 
one dimensional quasiperiodic lattices 
in particular, have occupied a large portion
of all these works [see 1-11, for example]. Perhaps the most widely
studied model is based on the Fibonacci sequence which started drawing interest
after the publication of a 
couple of pioneering  papers by Kohmoto et al [1] and Ostlund et al [2].
The spectral properties of one dimensional Fibonacci chain 
 are exotic. The single particle eigenstates 
are neither extended, nor localized in a conventional sense,   
and the spectrum is a Cantor set with zero Lebesgue measure [1-3].
The existence of self similarity and scaling behaviour of the wave functions
are two other major properties which mark the difference of 
such a quasicrystalline chain from
a periodic or a randomly disordered lattice.
A Fibonacci chain consists of two letters $A$ and $B$. The entire sequence
is generated by successive application of the substitution rule, 
$A\rightarrow AB$ and $B\rightarrow A$. The first few generations are,
$G_0 = A$, $G_1 = AB$, $G_3 = ABA$, $G_4 = ABAAB$, and so on, where the 
letter $G$ indicates `generation'.  
The
letters $A$ and $B$ may denote two different `atoms' (site model) or,
two different `bonds' separating identical atoms (transfer model) [7]. 
The electronic properties of such systems are well described within the
standard tight binding formalism, where the Hamiltonian, in the 
Wannier basis, is given by,
\begin{equation}
H = \sum_i\epsilon_{i}|i\rangle\langle i| +\sum_{ij}\left[t_{ij}|i\rangle
\langle j|+t_{ji}|j\rangle\langle i|\right]
\end {equation}
Here, $\epsilon_{i}$ is the site energy corresponding to the $i$-th site 
and $t_{ij}=t_{ji}$ is the hopping
integral between the $i-$th and the $j-$the sites. 
To solve for the eigenfunctions it is useful to find the
solutions of a set of difference equations which, in the nearest 
neighbor approximation look like, 
\begin{equation}
(E-\epsilon_i) \psi_i = t_{i,i+1} \psi_{i+1} + t_{i,i-1} \psi_{i-1}
\end{equation}
Here, $\psi_i$ is the amplitude of the wave function on the $i-$th site.
The above equation can equivalently be cast into a matrix form:
\begin{eqnarray}
\left(\begin{array}{c}\psi_{i+1}\\ \psi_i  
\end{array} \right)
& = & 
\left( \begin{array}{cc}
\frac{(E-\epsilon_{i})}{t_{i,i+1}}&-\frac{t_{i,i-1}}{t_{i,i+1}}\\1&0
\end{array} \right)
\left(\begin{array}{c}\psi_{i}\\ \psi_{i-1} 
\end{array} \right)
\end{eqnarray}

\noindent
The matrix, 
\begin{displaymath}
M_i=\left( \begin{array}{cc}
\frac{(E-\epsilon_{i})}{t_{i,i+1}}&-\frac{t_{i,i-1}}{t_{i,i+1}}\\1&0
\end{array}\right)
\end{displaymath}
which relates the amplitudes of the wave function at three consecutive
sites on the chain, is called the transfer matrix.

\noindent
For investigating the spectral properties of a quasiperiodic chain, one major 
 approach, using a method of the dynamical systems analysis, has been
to generate first a finite segment (a `rational approximant' [1,2,7]) 
of the Fibonacci chain following the
substitution rule. The infinite chain is then constructed by periodically
repeating this segment. In this way one can use higher and
higher order rational approximants to attain a true quasiperiodic Fibonacci
chain in the thermodynamic limit.  
For any generation $l$ one can construct a transfer matrix $M_l$ which is the 
product of the individual transfer matrices across each atomic site in 
that $l-$th generation, taken in proper order, i.e., $M_l=\prod_iM_i$.
This product matrix $M_l$ plays 
a vital role in determining the character of the spectrum and the 
eigenfunctions.
The allowed eigenvalues for the infinite sequence
are extracted from a study of the evolution of the trace of the above
transfer matrix at successive generations of the Fibonacci sequence [1,2,7]. 
Two important properties of this product matrix for any generation 
are now well established. We refer to them with respect to the golden mean
Fibonacci chain, which will play a central part in this communication.
\vskip .2in
\noindent
{\it (i) The Trace Map:} Following the prescription that generates the
sequence, it is easy to see that the
total transfer matrix for any $l+1$-th generation sequence is related to
the matrices corresponding to two previous generations by the relation, 
$M_{l+1}=M_{l-1}M_{l}$. This leads to a recursion relation between the traces 
of $M_n$, viz, 
\begin{equation}
x_{l+1}=2 x_l x_{l-1}-x_{l-2}
\end{equation}
Here $x_l=Tr(M_l)/2$.
The ``allowed" energy values are those for which $|x_l|\le 1$. As the generation
increases we find that almost all the energy values 
belong to the {\it escaping orbits}
of the map [1,2], which speaks for a true singular continuous  
character of the spectrum. 
\vskip .3in
\noindent
{\it (ii) The Matrix Map:} In addition to the above property, 
the product matrix $M_l$ at any generation $l$ is found to exhibit a 
cyclic behavior [1-7], viz, $M_{l+m}=M_l$ for certain energy eigenvalues that
depend on the model considered. To clarify, let us give the example of the
so called `transfer model' of the 
standard golden mean Fibonacci chain, where the letters $A$ and $B$ 
stand for two bonds, `Long' $(L)$ and `Short' $(S)$ respectively. All the
on site potentials are assumed to be same, and put equal to zero. 
In this model one finds that,  at $E=0$ which happens to be the
centre of the spectrum in this case, the matrix product 
$M_l$ has a {\it six cycle}, viz, $M_{l+6}=M_l$ for $l\ge 2$. 
It may be noticed that such cyclic
behaviour of the transfer matrix is also common to quasiperiodic chains
other than the golden mean [10]. It is this cyclic property 
of the matrix $M_l$, that has stimulated
the present work. The motivation will be clear from the 
discussion that follows. 

\vskip .3in
\noindent
The cyclic behaviour of the transfer matrices, as it appears from 
the previous works in general, has been shown to exist  
for various quasiperiodic chains without any underlying physical
reason. It is not straightforward to prove the existence of
cycles of the matrix map for any arbitrary substitutionally generated
sequence. Not only that, even if it exists, it is extremely difficult to
locate the precise energy eigenvalue for a lattice such as the 
Fibonacci chain in which the spectrum is a  Cantor set 
having a gap
at the vicinity of every energy, at which such a cycle will take place.
Precisely, this is the point that motivates us to undertake the present
study. We wish to examine whether it is at all possible to provide a simple
physical picture  that will lead to an understanding the origin of 
cycles of the matrix map
based on the geometric structure of the quasiperiodic chain concerned.
In this communication we begin by examining the 
well known six cycle behaviour at $E=0$ 
in the simplest transfer model
of a golden mean Fibonacci chain. We show that there exists hidden {\it dimer}-
like correlations between the constituent `atoms'. The dimers are not
explicitly displayed, but their presence can be directly linked with the
six cycle of the full matrix map at $E=0$. 
We give a prescription to unravel
the other six cycle energy values as well, though, 
$E=0$ is the only energy for which we have been able to detect a six cycle of the matrix map in the present model.  
However,
in a mixed model of the Fibonacci chain we show that following the same
prescription we can evaluate at least {\it two} 
values of the electron energy at
which the matrices will repeat after every six generations. 
Our method is perfectly general, and can be applied to any quasiperiodic
sequence. In the latter part of this paper we extend the study
to the quasiperiodic silver mean chain [10] to show the presence of similar
kinds of dimers and establish their relationship with the 
{\it four-cycle} matrix
map at $E=0$ already reported in the literature [10].
\vskip .3in
\noindent
In this context it is worth mentioning that the existence of dimers in
random and quasiperiodic chains [13-16] has been known to cause unscatterd/
extended eigenstates leading to a perfect transmission. However, the
Fibonacci chain and all such lattices in the golden mean class,
generated for example, using the rule $A \rightarrow A^nB$, $B \rightarrow A$
[10] 
have so far remained un-noticed in this respect.
In a very recent
work on a more involved model of the Fibonacci lattice 
Satija et al [16] have shown that
there exist a single `resonant' eigenstate that 
can be explained in terms of the 
presence of a `dimer'
We begin by studying  the oldest and the simplest version of the transfer model
of a Fibonacci chain.
It will be shown  that the dimers in this system 
exist naturally, but they combine
in a nested fashion to give rise to a `resonance' 
only when certain conditions are satisfied.
In what follows we describe the method of analysis.
\vskip .3in
\noindent
The transfer model of a Fibonacci chain comprises
of two different `bonds' $L$ and $S$ generated according to the prescription
$L\rightarrow LS$ and $S\rightarrow L$. 
We attribute three values to the on site potentials,
viz,
$\epsilon_{\alpha}$, $\epsilon_{\beta}$ and ,
$\epsilon_{\gamma}$ corresponding to the sites flanked by $L-L$ ,
$L-S$, $S-L$ 
bonds respectively. The nearest neighbor hopping integrals are designated
by $t_L$ and $t_S$ corresponding to the $L$ and $S$ bonds [9].  
Let us now define three types of transfer 
matrices correspondent to the three sites $\alpha$, $\beta$ and $\gamma$, viz
\begin{displaymath}
M_\alpha=\left( \begin{array}{cc}
\frac{(E-\epsilon_{\alpha})}{t_L}&-1\\1&0
\end{array}\right);
M_{\beta}=\left( \begin{array}{cc}
\frac{(E-\epsilon_{\beta})}{t_S}&-\frac{t_L}{t_S}\\1&0
\end{array}\right); 
M_{\gamma}=\left( \begin{array}{cc}
\frac{(E-\epsilon_{\gamma})}{t_L}&-\frac{t_S}{t_L}\\1&0
\end{array}\right) 
\end{displaymath}
The transfer matrix for any $l+1$-th generation chain is obtained recursively
following the equation,
\begin{equation}
M_{l+1}=M_{l-1}M_l
\end{equation}
where, $M_l$ is the matrix for the $l$-th generation, obtained by multiplying
the above elementary matrices $M_\alpha$, $M_\beta$ and $M_\gamma$  
in proper order. 
In the above equation, $M_1=M_\alpha$ and $M_2=M_{\gamma}M_\beta=
M_{\gamma\beta}$ (say) [7]. 
It has been observed in the literature [7] that in the above model with
$\epsilon_\alpha=\epsilon_\beta=\epsilon_\gamma=0$ and $t_L=1$ and $t_S=2$,
one gets $M_{l+6}=M_l$ for $l\ge 2$. Let us now give a simple
physical picture of this phenomenon based on the existence of the dimer like
positional correlations in this model of the Fibonacci chain.
\vskip .3in
\noindent
Consider the second generation chain, i.e. $l=2$. This is $LS$. 
The product transfer matrix,
when written explicitly for this segment, is, 
$M_2=M_{\gamma}M_{\beta}M_{\alpha}$. We have taken the first site of this chain
to be of type $\alpha$ without the loss of generality. 
The last site will be of type $\gamma$. This is obvious, because the
segment $LS$ has to be repeated periodically, 
so that the bond appearing after the $S$ will be $L$. The
matrix for $l=8$ i.e. $M_8$ naturally consists of a much longer string
of $M_\alpha$, $M_\beta$ and $M_\gamma$. Yet, at $E=0$, we all know that
$M_8=M_2$ and this is repeated after every six steps for $l\ge 2$. 
So, to our mind, a natural
question should be, how do the individual matrices combine so that the
bigger and bigger strings of matrices shrink back to $M_2$ at this
particular energy ? This leads us to carefully examine the structure of
the Fibonacci chain.
In Fig.1 we show a Fibonacci chain in the sixth generation for example. 
We observe
that in it (as well as in an infinite chain), 
the $\alpha$-sites appear in 
isolation and the $\beta-\gamma$ sites appear in pair, as well as in 
isolation. It is important to observe that every 
isolated $\beta-\gamma$ cluster, as well as $\beta-\gamma$ pairs side by side, 
are flanked by two 
$\alpha$-sites. This happens 
everywhere,  even in the infinite chain. 
Now, if the trace of the matrix
$M_{\gamma\beta}=M_{\gamma}M_{\beta}$ becomes zero, 
then the corresponding energy renders the
matrix product $M_{\gamma\beta}M_{\gamma\beta}$ identity (barring a negative
sign) [12-14]. 
Once the $\beta-\gamma$ clusters (joined by the solid curve in Fig.1) 
combine to give us an identity contribution in terms of the matrices,
the $\alpha$ sites at the two flanks (joined by a dotted curve in Fig.1)
automatically form a `dimer' [12]. 
If the trace of $M_\alpha$ now vanishes for the {\it same energy}, 
then the product of two consecutive $M_\alpha$ matrices also become equal to
an identity matrix (once again with a negative sign), 
so that a local cluster of matrices 
$M_{\alpha}M_{\gamma\beta}M_{\gamma\beta}M_\alpha$ becomes equal to an 
identity matrix. This will happen locally {\it throughout the chain}. 
In this way one can see that the entire string of matrices will 
reduce, in `size',  because of the resonance in these 
two different kinds of dimers ($\alpha-\alpha$ and $\beta\gamma-\beta\gamma$)
simultaneously.
These pairs appear in a nested fashion, and appear at all scales of length
because of the self-similar character of the lattice. 
If we get an energy for which $M_\alpha^2=M_{\gamma\beta}^2=-I$,
$I$ being the $2X2$ identity matrix, 
 then one can verify that,
$M_l=M_{l+6}$ for $l \ge 2$. Now, let us see if $E=0$ satisfies this
criterion. 
In the transfer model,   
\begin{equation}
Tr(M_{\gamma\beta})=\frac{(E-\epsilon_\beta)(E-\epsilon_\gamma)-
(t_L^2+t_S^2)}{t_Lt_S}
\end{equation}
We select $\epsilon_\alpha=\epsilon_\beta=\epsilon_\gamma=0$ and $t_S=2$ and
$t_L=1$ [7]. $E=0$ definitely makes $M_\alpha^2=-I$, but $Tr(M_{\gamma\beta})
 \ne 0$ and hence, $M_{\gamma\beta}^2 \ne I$. The resonance condition
discussed above is not satisfied. 
This of course, does not rule out the possibility of having
bigger clusters in the formation of nested dimers. 
For example, one can see (Fig.2a) that there are clusters formed by the
triplet $\beta\gamma\alpha$ and $\beta\gamma$ which are distributed in the
same manner and can be considered to be a `renormalized' version of the
basic clusters formed by $\beta\gamma$ and $\alpha$ respectively. It turns
out that in the transfer model  
the clusters responsible for a six cycle of the matrix map at 
$E=0$ are the pairs $\beta\gamma\alpha\beta\gamma$-
$\beta\gamma\alpha\beta\gamma$ flanked at the
two extremities by the cluster $\alpha\beta\gamma$. That is,
$Tr(M_{\gamma\beta}M_{\alpha}M_{\gamma\beta})=Tr(M_{\gamma\beta}M_\alpha)=0$ 
at $E=0$. Under this condition $M_{l+6}=M_l$ for $l \ge 2$.
Once this is observed the question that arises immediately is, how to
identify the {\it minimal} clusters responsible for a possible six (or other)
cycle of the full matrix map for the chain. To resolve this issue we
get back to the method of real space renormalization scheme, applied to
the infinite quasiperiodic chains in the literature [9]. It is important
however, to appreciate that, the renormalization group method will be used
here only to identify the minimal clusters responsible for resonance.
\vskip .3in
\noindent
It is known that the on-site potentials and the hopping integrals in the 
infinite Fibonacci chain renormalize to [9]: 
\vskip .2in
\noindent
$\epsilon_\alpha(n+1)=
\epsilon_\gamma(n)+[t_L(n)^2+t_S(n)^2]/(E-\epsilon_\beta(n));
\epsilon_\beta(n+1)=\epsilon_\gamma(n)+t_S(n)^2/(E-\epsilon_\beta(n));\\
\epsilon_\gamma(n+1)^2=\epsilon_\alpha(n)+t_L(n)^2/(E-\epsilon_\beta(n));
t_L(n+1)=t_L(n)t_S(n)/(E-\epsilon_\beta(n))$ and $t_S(n+1)=t_L(n)$.
\vskip .2in
\noindent
One defines a  parameter space of reduced dimensionality, and
defined by the dimensionless quantities $w,x,y$ and $z$, where,
\vskip .2in
$w_n=(E-\epsilon_{\alpha}(n))/t_{L}(n),
x_n=(E-\epsilon_{\beta}(n))/t_{L}(n),
y_n=(E-\epsilon_{\gamma}(n))/t_{L}(n)$ and, 
$z_n=t_{S}(n)/t_{L}(n)$.
These parameters, under renormalization give rise to the following set
of recursion relations [9],

\noindent
$w_{n+1} = (x_n y_n)/z_n -z_n -1/z_n   ;
x_{n+1} = (x_n y_n)/z_n -z_n   ;
y_{n+1} = (w_n x_n)/z_n - 1/z_n$, and, 
$z_{n+1} = x_n/z_n$.

\noindent
From the above recursion relations one can show that 
\begin{equation}
w_{n+3}=w_{n+2}w_{n+1}-w_n
\end{equation} 
which is equivalent to the trace map $(3)$ with $w_n=2 x_n$.
Using the set of recursion relations (7) together
with Eq.(6) it is straightforward to show that if we set $w_n=w_{n+1}=0$
then 
{\it (i)} $w_{n+3}=w_n$, 
and {\it (ii)} $(w_{n+6},x_{n+6},y_{n+6},z_{n+6})=(w_n,x_n,y_n,z_n)$. 
That is, the entire parameter space is {\it mapped onto itself} 
after six steps of
renormalization. Additionally, one comes across a three cycle of the
trace map (as $w_{n+3}=w_n$). 
As the total transfer matrix across any arbitrary generation
can be completely expressed in terms of the parameters $w$, $x$, $y$ and $z$, 
it is obvious that the transfer matrix for the generation $n+6$ will be
identical to that in the generation $n$ for the energy value extracted by
setting $w_n=w_{n+1}=0$. Most interestingly, it can be checked 
that $w_n$ is actually $Tr(M_{\alpha,n})$ and $w_{n+1}=
Tr(M_{\alpha,n+1})=Tr(M_{\gamma\beta,n})$
, where $M_{i,n}$ is the transfer matrix for the $i-$th site 
in the $n-$ step renormalized lattice. Therefore, setting $w_n=w_{n+1}=0$
is equivalent to searching for {\it simultaneous zeros} 
for the polynomial equations
$Tr(M_{\alpha,n})=0$ and $Tr(M_{\gamma\beta,n})=0$. 
If we start with an infinite Fibonacci chain and re-scale it, 
we observe that the resonance condition, as described above,
is satisfied for the first time on a two times renormalized version of the
original Fibonacci chain (i.e. $n=2$) if we take $E=0$. 
That is, for $E=0$, we get $w_2=0$ and $w_3=0$ simultaneously.
The $\alpha-\alpha$ and the $\gamma\beta-\gamma\beta$ pairs at this 
length scale form the required `dimers' causing resonance. 
To see what minimal clusters do they correspond to in terms
of the original chain, we trace back
to the lattice at the original length scale to see that the elementary
blocks causing resonance are clusters like $\alpha\beta\gamma$ and 
$\beta\gamma\alpha\beta\gamma$ respectively. 
The `dimers' in a one-step
renormalized lattice and their correspondence with 
the `dimers' in the original, un-renormalized lattice is 
illustrated in Fig.2. Doing
one more step is simple. 
It should be noted that the method outlined here
indicates the possible existence of six cycle for other values of $E$. 
For example, one could search for simultaneous zero's of, say, 
$w_4$ and $w_5$ etc. However,
we have been unable to detect any other value for the energy in this model
that may lead to other cyclic behaviors.
But, we propose a second model which clearly shows that 
based on the same thread of reasoning one can get more
than one energy values for which a six cycle map for the transfer matrix
exists.
\vskip .3in
\noindent
{\it A mixed model:}
\vskip .2in
\noindent
Let us select a model in which $\epsilon_\alpha=\sqrt{5}$, $\epsilon_\beta=
\epsilon_\gamma=0$, $t_L=1$ and $t_S=2$. With this choice of the
parameters we find that, $Tr(M_\alpha)=Tr(M_{\gamma\beta})=0$ in the 
original (un-renormalized) lattice
if we set
$E=\sqrt{5}$. This leads to the fact that if we start with the above
mixed model Fibonacci chain at the $8$-th generation, then the
$\alpha-\alpha$ and $\beta\gamma-\beta\gamma$ pairs in the original
lattice  
will form the nested dimers. The corresponding transfer matrices will 
combine together to give rise to an identity matrix, and one is finally
left with a product of $M_{\gamma\beta}M_\alpha$ in the $8-$th generation
corresponding to the triplet of sites $\alpha\beta\gamma$ which is
the same as the second generation chain. 
That is, we achieve the result $M_8=M_2$, 
$M_9=M_3$ and so on.
Let us
now renormalize the lattice once and 
look for a simultaneous solution of the 
equations $w_2=0$ and $w_3=0$. Most interestingly we find that $w_2=w_3=0$ 
for $E=-\sqrt{5}$. If we trace back to the original lattice it is easy to
discern that the fundamental clusters responsible for resonance in this case
are $\alpha\beta\gamma$ and $\beta\gamma$ respectively. 
This implies that for $E=-\sqrt{5}$ the six cycle feature 
$M_l=M_{l+6}$ sets in for $l\ge 3$.
Thus
we have been able to identify two different energy eigenvalues for this
mixed model of the Fibonacci chain for which the matrix map has a six cycle.
For the two different energy eigenvalues the `dimer correlations' are
revealed at {\it two different scales of length}. This is an important feature, 
linked with the inherent self similarity of the lattice concerned.

\vskip .3in
\noindent
{\it The silver mean lattice:}
\vskip .2in
\noindent
To check the validity of our arguments further, we have extended this idea
to the transfer model of the  silver mean lattice, 
which is generated according to the
rule $L\rightarrow LLS$ and $S\rightarrow L$. It may be mentioned at this
point, that the entire discussion made for the silver mean case is
valid, in general, for the class of lattices grown according to the
rule $L \rightarrow L^mS$, $S \rightarrow L$.
The recursion relations depicting the parameter
space in the silver mean case are [17],

\noindent
$w_{n+1} = (w_n x_n-1)y_n/z_n-w_n z_n-x_n/z_n   ;
x_{n+1} = (w_n x_n-1)y_n/z_n-w_n z_n   ;
y_{n+1} = (w_n x_n-1)w_n/z_n -x_n/z_n   ;
z_{n+1} = (w_n x_n-1)/z_n$, and, 
$t_{n+1} = x_n y_n/z_n -z_n -1/z_n$

\noindent
From the above relations it follows that [17],
$w_{n+2} = w_{n+1}t_{n+2} - w_n$ and, $t_{n+2} = w_nw_{n+1} - t_{n+1}$.
\noindent
The basic clusters responsible for dimer correlation
in this case are $\alpha-\alpha$ and $\alpha\beta\gamma-\alpha\beta\gamma$
respectively. We find that, $(w_{n+4},x_{n+4},y_{n+4},z_{n+4})=
(w_n,x_n,y_n,z_n)$ if we set $w_n=w_{n+1}=0$. This happens for $E=0$.
That means the four cycle of the matrix map [10] is observed following
our prescription. The trace map exhibits a two cycle behavior in this case,
as $w_{n+2}=w_n$.
One can look for other possible cycles of $M_l$ by renormalizing the
lattice and following the arguments given earlier. These aspects are
now being investigated and the results will be published in due
course.
\vskip .3in
\noindent
{\bf Acknowledgment:}
\vskip .2in
\noindent
The authors are thankful to R. K. Moitra and S. N. Karmakar for suggestions.
One of the authors (A.C.) acknowledges interesting discussions with
Michael Schreiber.
\pagebreak

{\bf References}
\vskip .3in
\noindent
[1] M. Kohomoto, L. P. Kadanoff and C. Tang, Phys. Rev. Lett. {\bf 50}, 1870
 (1983). 

\noindent 
[2] S. Ostlund et al., Phys. Rev. Lett. {\bf 50}, 1873 (1983). 

\noindent
[3] M. Kohomoto and Y. Oono, Phys. Lett. A {\bf 102}, 145 (1984); M. Kohmoto
and J. R. Banavar, Phys. Rev. {\bf B 34}, 563 (1986); J. P. Lu, T. Odagaki
 and J. L. Birman, Phys. Rev. {\bf B 33}, 4809 (1986); F. Nori and
J. P. Rodriguez, Phys. Rev. {\bf B 34}, 2207 (1986).

\noindent
[4] 
B. Simon, Adv. Appl. Math. {\bf 3}, 463 (1982); J. B. Sokoloff, 
Phys. Rep. {\bf 126}, 189 (1985); J. M. Luck and D. Petritis, J. Stat. Phys.
 {\bf 42}, 289 (1986).

\noindent
[5]. C. Tang and M. Kohmoto, Phys. Rev. {\bf B 34}, 2041 (1986);
B. Sutherland, Phys. Rev. Lett. {\bf 57}, 770 (1986);

\noindent
[6] F. Deylon and D. Petritis, Commun. Math. Phys. {\bf 103}, 441 (1986);
Q. Niu and F. Nori, Phys. Rev. Lett. {\bf 57}, 2057 (1986).

\noindent
[7] M. Kohmoto, B. Sutherland and C. Tang, Phys. Rev. {\bf B 35}, 1020 (1987).

\noindent
[8] Y. Liu and R. Riklund, Phys. Rev. {\bf B 35}, 6034 (1987); R. Riklund, 
M. Severin and Y. Liu, Int. J. Mod. Phys. {\bf B 1}, 121 (1987).

\noindent
[9] J. A. Ashraff and R. B. Stinchcombe, Phys. Rev. {\bf B 37}, 5723 (1988).

\noindent
[10] G. Gumbs and M. K. Ali, Phys. Rev. Lett. {\bf 60}, 1081 (1988);
G. Gumbs and M. K. Ali, J. Phys. {\bf A:}Math. Gen. {\bf 22}, 951 (1989);
Q. Niu and F. Nori, Phys. Rev. {\bf B 42}, 10329 (1990);
J. M. Luck, Phys. Rev. {\bf B 39}, 5834 (1989); A. Bovier, J. Phys. {\bf A}:
Math.Gen. {\bf 25}, 1021 (1992); A. Bovier and J. M. Ghes, Commun. Math. 
Phys. {\bf 158} 45 (1993).

\noindent
[11] A. Sanchez, E. Macia and F. Dominguez-Adame, Phys. Rev. {\bf B 49}, 
147 (1994); E. Macia and F. Dominguez-Adame, Phys. Rev. Lett. {\bf 76}, 
2957 (1996); {\it ibid} {\bf 79}, 5301 (1997).

\noindent
[12] E. Macia and F. Dominguez-Adame, in {\it Electrons, Phonons and
Excitons in low dimensional aperiodic systems}, and references therein,
Editorial Coplutense, Madrid (2000). 
 
\noindent
[13] D. H. Dunlap, H-L. Wu and P. W. Phillips, Phys. Rev. Lett. {\bf 65},
88 (1990).

\noindent
[14] Arunava Chakrabarti, S. N. Karmakar and R. K. Moitra, 
Phys. Rev. {\bf B 50}, 13276 (1994).

\noindent
[15] Arunava Chakrabarti, S. N. Karmakar and R. K. Moitra, 
Phys. Rev. Lett. {\bf 74}, 1403 (1995).

\noindent
[16] I. G. Guesta and Indubala Satija, {\it Cond-mat/9904022}, {\bf April 1}
(1999).

\noindent
[17] Arunava Chakrabarti and S. N. Karmakar, Phys. Rev. {\bf B 44}, 896 (1991).
\newpage
\noindent
{\bf Figure Captions:}
\vskip .3in
\noindent
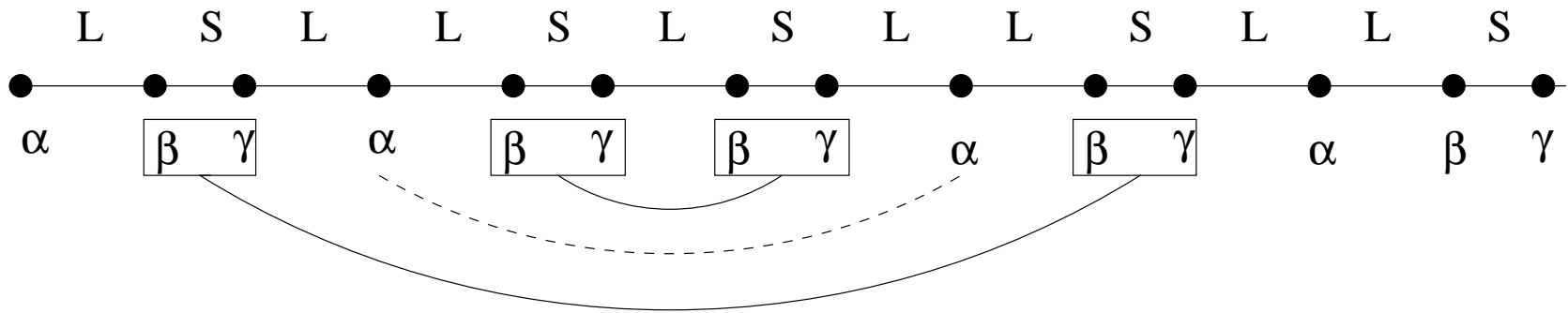
{\bf Fig.1:} Sixth generation Fibonacci chain with three different
sites. Clusters forming dimers are displayed by solid (for $\beta-\gamma$)
and dashed (for $\alpha$) lines respectively.
\vskip .3in
\noindent
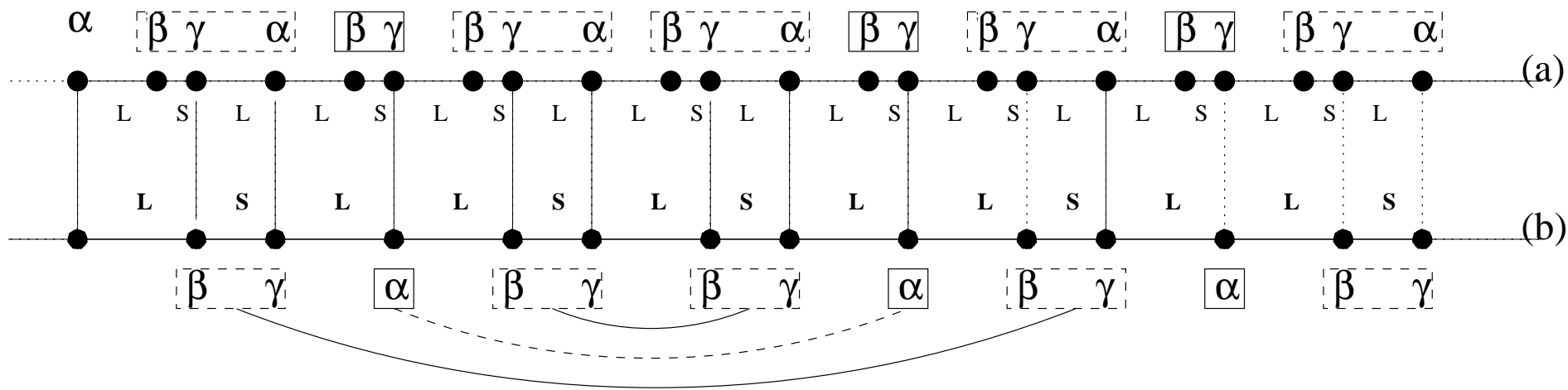
{\bf Fig.2:} Portion of an infinite Fibonacci chain.  
Clusters responsible for dimer-like correlations in 
one-step renormalized lattice (b) are
the $\beta-\gamma$ doublet (dashed box) and the single
$\alpha$ sites (solid box). They correspond to the triplet $\beta\gamma\alpha$
(dashed box) and the doublet $\beta\gamma$ (solid box) 
respectively,in the un-renormalized
lattice (a). 
\end{document}